\documentclass[11pt,aps,amsmath,amssymb,nofootinbib,notitlepage,longbibliography]{revtex4-1}
\usepackage{amsmath,amssymb}
\usepackage{slashed}
\usepackage{graphicx}
\usepackage{bm}
\usepackage[top=1.0in,bottom=1.0in,left=1.0in,right=1.0in]{geometry}
\usepackage[colorlinks,linkcolor=blue,citecolor=blue]{hyperref}

\newcommand{\be}{\begin{equation}}
\newcommand{\ee}{\end{equation}}
\newcommand{\bea}{\begin{eqnarray}}
\newcommand{\eea}{\end{eqnarray}}
\newcommand{\ba}{\begin{eqnarray}}
\newcommand{\ea}{\end{eqnarray}}

\begin{document}

\title{ The heavy ions are ``preheated" prior to high energy collisions}

\author{ Edward Shuryak}
\email{ Edward.Shuryak@stonybrook.edu}

\affiliation{Center for Nuclear Theory, Department of Physics and Astronomy, Stony Brook University, Stony Brook, New York 11794--3800, USA}

\begin{abstract}
The so called ``isobar run" or RHIC was designed to compare a number of observables for collisions of $^{96}_{40}Zr$ with those of  $^{96}_{44}Ru$, aimed at identification of $Z$-dependent effects. However, as the STAR data have shown with unprecedented accuracy,
these two nuclides  differ stronger than expected, producing effects larger than those depending on charge.
So far, multiple studies tried
to quantify their shape differences, in relation to various observables. General consensus is that these
differences somehow should 
be related to nuclear structure, in particularly properties of the lowest excited states. Yet the precise connection
between these fields -- low and high energy nuclear physics -- is still missing. In this paper I propose such a connection,
via a concept of  thermal density matrices of a ``preheated" nuclei. The effective temperature should 
parameterize which set of excited states should be included in the calculations. I also 
suggest  semiclassical ``flucton" method at finite temperatures to be used to calculate thermal density matrices.
\end{abstract}

\maketitle
\section{Introduction}

Specific selection of $^{96}_{40}Zr$ and  $^{96}_{44}Ru$ for RHIC run
was based on the original idea  that the same total number of nucleons will make   backgrounds very similar in two cases, and the difference
would be mostly related to different electric charges, revealing
in particular Chiral Magnetic Effect (CME).
However,  the resulting data set \cite{STAR:2021mii} has shown that it is {\em not the case}: observables like multiplicity distribution,
elliptic and triangular flows and many similar flow observables show nontrivial
differences between these two nuclei.  The non-electromagnetic background effects 
turns out to be of the order of several percents, comparable or larger than expected CME effect. 
On the other hand, from experimental point of view this ``RHIC isobar run" is very successful, since the accuracy of measurements reached is unprecedented $\sim 0.4 \%$, smaller than differences between $Ru$ and $Zr$. These data
give us an opportunity to test better current models of heavy ion collisions.

So, {\em how different} may two nuclides used, $^{96}_{40}Zr$ and  $^{96}_{44}Ru$,  be? 
Just four of  neutrons
are turned to protons, so all effects should be proportional to  small factor $4/96$. The second factor which enters is  the {\em difference} between  states in which these protons/neutrons are located. 

Naive ``liquid drop" model would suggest that these differences are due to Coulombic  
 repulsion, pushing $protons$ to larger radii. As we will soon see, it is not true.
 For these nuclei Coulomb potential is in fact
small (compared to nuclear ones), so this effect is also
to rather small, at a sub-percent level.

\begin{figure}[t!]
\begin{center}
\includegraphics[width=8cm]{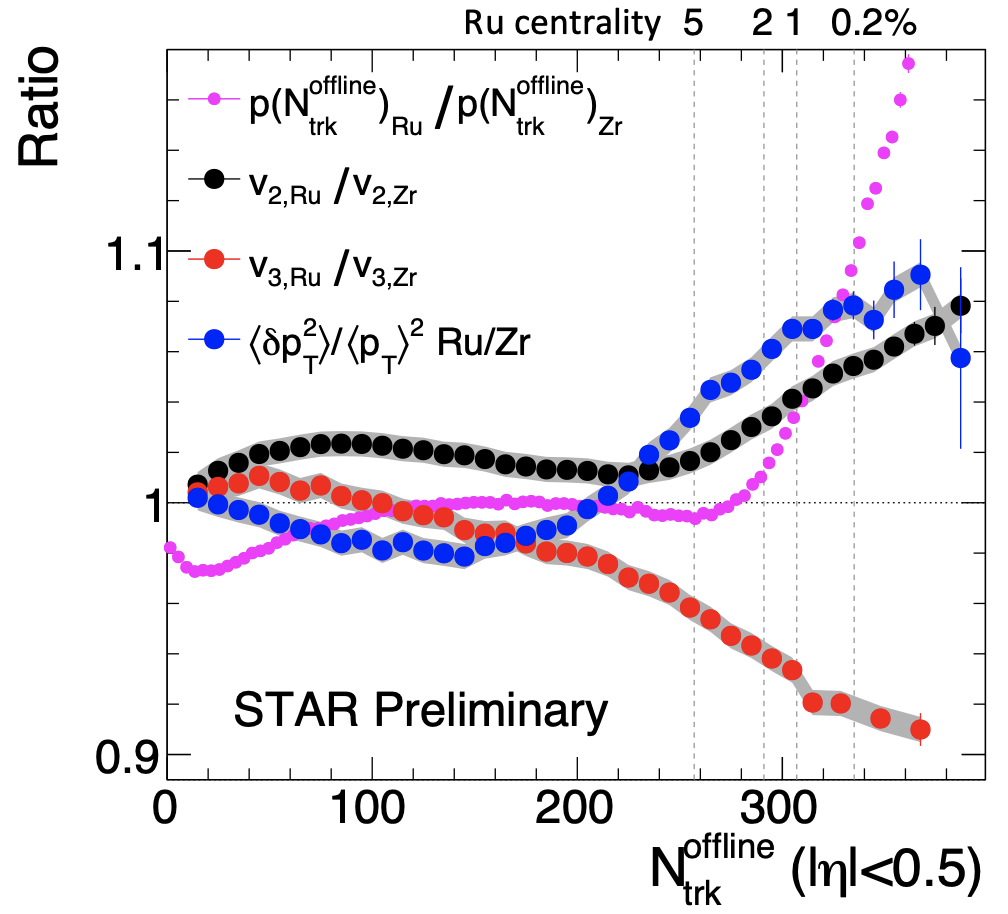}
\caption{Four observables defined in the plot, all in form of ratios for $Ru Ru$ to $Zr Zr$
collisions from STAR as a function of centrality. The lower scale is number of tracks per unit rapidity, the upper scale with thin dashed lines are percentage of  total cross section.}
\label{fig_STAR_ratios}
\end{center}
\end{figure}

A compilation of the ratios for 4 observables measured by STAR collaboration (from \cite{2109.01631}) is shown in Fig.\ref{fig_STAR_ratios}.
Apparently, for noncentral collisions (the left side of the plot, $>5\%$) deviations are
at $2\%$ level,  but for ultra-central collisions they reach the level of $O(10\%)$ or more.
 Thus,  
 naive original expectations were incorrect, and these two nuclei turned out to be in fact very different!
 

In order to understand {\em why} $^{96}_{40}Zr$ and  $^{96}_{44}Ru$ are 
so different,   and be able to quantitatively explain these results, one needs to turn to 
  nuclear structure. In particular, we will discuss:\\
(i) the wave functions of ``valence" quasiparticles, $p$ ``holes" and $n$ ``particles" in nuclear shell model,
see Appendix  \\
(ii) structure of ``excitation trees" for both nuclei,  which can be understood as ``vibrators" and ``rotators"
with certain parameters to be used in description of 
shape fluctuations. 
Rotational bands indicate deformation of $Ru$, but not $Zr$.\\
(iii) the range of energies $\Delta E$ out of which one need to take excitations to reproduce the shape of the
virtual state 
in which one finds nuclei at the collision moment. We will interpret it via ``preheating temperature" parameter $T_\perp$.
To introduce  the width of this range is  $unavoidable$ because nuclei are not rigid objects. Moments of inertia
and nondiagonal matrix elements among the first couple of levels and the first 20 or so are {\em not the same}.
So, to quantify fluctuations of nuclear shapes, one needs to define which set of excitations are involved.

\section{History of the notion of ``intrinsic nuclear shape" and its manifestations in heavy ion collisions} 
The idea that some nuclei are spherical and some are deformed goes back to 1950's and
by now it takes its proper place in textbooks. What is important for this paper is to underline is that
it is $not$ formulated in terms of the ground state wave function $|0\rangle$  (which for even-even nuclei
are ``spherical" $0^+$ always). 

What is very important is that  these theories aim not only at description of  the ground state but also certain number of {\em excitations}. 
Certain sets of these states $|n \rangle$ are
interpreted in terms of particular models, such as ``rotator", "vibrator" etc. Their properties are: the  momentum of inertia $\Theta$
in the former case, vibrational frequency in the latter  and  transition matrix elements. So, excitation energies $E_n$, diagonal and non-diagonal matrix elements of operators (such as magnetic dipole or electric quadrupole  moments) are 
all involved when one describes ``nuclear shapes" via
 projections onto dynamics of certain
``collective variables" describing deformation $ \beta_2,\gamma, \beta_3...$.

These variables do not  possess some fixed ``classical" values.
Virtual states possessing multiple values of collective variables
$| \beta_2,\gamma , ...\rangle $ 
are then put in form of model Hamiltonians containing certain ``potentials"  
\be \langle  \beta_2,\gamma , ..| \hat H | \beta_2,\gamma , ...\rangle=E(\beta_2,\gamma...) 
\ee
to be used to explain quantum dynamics of the nuclei.

For example, if the minimum happens to be at zero variables $\beta_2=\beta_3=0$, the nuclei is ``spherical". 
But of course, there are quantum oscillations around the minimum, which we
describe as ``phonon" states.  If these potentials can have two or more
minima, one may define several  ``vibrators" and look for their excitations among the experimentally observed states.
Sometimes the potentials are flat in a wide range of variables, and the nucleus is declared ``soft" with indefinite shape.
We will return to specific examples of that relevant for our "isobar" nuclei below, from nuclear structure literature.

Not being involved in any of that, I first met the issues we will discuss in this paper in the last year of the previous century
(just before the first run of RHIC) 
\cite{Shuryak:1999by} considering what would happen if we would collide a well-deformed nucleus such as $^{238}U$. 
It seemed obvious that classical notion of random directions of deformation axes of both nuclei would lead to
a variety of situations (``tip-to-tip" etc), and my then primitive simulations addressed a question whether one be
able to distinguished them experimentally. Sending the paper to PRC I got a referee report proclaiming
the paper wrong and very misleading. The argument was that since the ground state is $J^P=0^+$, it is spherically
symmetric, and thus the idea of intrinsic nuclear shape is nonsense.  

My defense was the argument  that I actually meant not the ground state but a wave packet made out of  $many$ excited states.
(The same idea as in this note.)
It eventually succeeded and paper  get  published (but it
took time, moving publication of the paper into the next millennium) . 

Yet then I would have hard time to
explain which specific set of excitations one would need to use. This is the question entertained in the present note.
We will suggest a very direct -- albeit still  model-dependent -- way to use the potentials $E(\beta_2,\gamma...)$ from
nuclear structure calculations to define the initial state in high energy heavy ion collisions.  

\section{Measurements and the density matrices} 
But before we discuss the main issue here, let me mention a previous problem I was involved with,
that of nuclear clustering and their influence on light nuclei production. Imagine several nucleons
forming some ``precluster", which at the ``freezeout moment" would go free into physical final states.
Here we have some virtual wave package being ``measured", namely get decomposed into states of the Hamiltonian.

Specifically, we  \cite{Escobar-Ruiz:2016aqv} discussed the problem of how a cluster of four nucleons can go into states of $He^4, He^3+p,t+n, d+d, ppnn$. 
Well, even not so many particles still have 12 coordinates, and working with 12 (or 9 of center of mass motion is eliminated)
dimensional function is not practical, so one need a single {\em collective variable}. Fortunately, it was known
to be a $hyperdistance$ -- the sum of all 9 Jacobi coordinates squared, which is just proportional to sum of all six
inter-particle distances  $\rho^2\sim \sum_{i>j} R^2_{ij}$. Therefore we set up a task to calculate
the corresponding {\em density matrix}, traced over all variables but $\rho$. In other words, 
we set up a calculation of the distribution over it, $P(\rho)$, in a matter made of interacting nucleons.

(The next step -- decomposition in Hamiltonian states -- is also fortunately simplifies, since
 it was shown already in 1960's that e.g. the ground state of $He^4$  is very well described by a
 function of this  single variable, the same hyperdistance $\rho$. In fact we even found it to be so
 for the $second$ excited state of $He^4$ as well.) 

Before going into description of technical tools used, let me emphasize a connection  of this problem to the problem at hand.
In both we try to establish a connection between a {\em set of stationary states} of the Hamiltonian with a {\em virtual state},
possessing certain distribution over some collective variables.   

Formally, one may think of this collective density matrix  to be calculated from  all Hamiltonian stationary states,  with a trace over all coordinates but special ones, taken with some  coefficients $ P_n$
 \be \label{eq:denmatr}
P(X)=\sum_{n=0}^\infty \int_{x_i} |\psi_n(X,x_i)|^2  P_n  
\ee
where summation over all non-collective coordinates $x_i$ is implied. 

In the 4-nucleon problem we had  a drastic simplification: the preclusters we were looking for came
from well equilibrated matter. Therefore we cold use Boltzmann factors as the proper weight $P_n=exp(-E_n/T)$.
If so, the collective distribution $P(X)$ is nothing else but a thermal density matrix, and the temperature is well measured freezeout temperature $T_f$.  As we will discuss soon,
there are multiple theoretical tools for its calculation available.

\section{``Preheating" of nuclei before the collision moment }
The act of high energy collision of two nuclei leads to ``act of measurements" of 
locations of the nucleons. We however prefer to describe those in terms of nuclear ``shape parameters",
localizing values of collective variables instead.

If there is no excitation, probability  $P(X)$ to find value $X$ is $|\psi_0(X)|^2$, based on
ground state wave function of the corresponding ``vibrator".  If there are high excitations,
classical thermodynamics suggest distribution to be Boltzmann $exp(-V(X)/T)$.
We will argue that the probability distribution over collective variables $P(\beta_2,\gamma)$ at the moment of a collision 
can be modeled by {\em quantum-thermal density matrix} which is in between these two limits.

%

Generically, an  act of measurement fixes nucleon transverse coordinates within certain 
uncertainty $\Delta \vec x_i$, resulting in
an uncertainty in the total energy $\Delta E$. Therefore not just
ground state byt excited ones,  from a strip $E_n < \Delta E$ would
 contribute to the density matrix (\ref{eq:denmatr}) of the virtual state. 
 
 Now, what the probabilities $P_n$ in that expression should be? Here we would like to evoke
 standard statistical argument. Suppose $\Delta E$ is large enough to encompass a large number of state
 which contribute about equally, so that
  the most important factor in the sum over states would be simply the density of states itself, or its entropy
 \be N(E)\sim exp[S(E)] \ee
 Standard expansion of it, with $\Delta S/\Delta E=1/T$, generates Boltzmann distribution over
 state's energies $exp(-E/T)$. 
 In other terms, one may think  that nuclei at the collision moment can be viewed as ``preheated" ones. 
 If so, the density
matrices over relevant collective variables can be evaluated as thermal ones.

Here comes the main question: what can this temperature of preheating be? Is it that we just
suggest that ``in anticipation" of the QGP production in the collision, the preheating temperature be what we 
usually call $T_0$ in hydrodynamical applications, namely hundreds of $MeV$s ?
The answer is $no$, it is not. The reason is equilibration of all degrees of freedom to a common $T_0$
require certain time, and is commonly assumed to be about $\tau\sim 1/2 \, fm/c $ $after$ the collision. 
At the collision moment one should discuss transverse and longitudinal degrees of freedom separately.

The accuracy of  localization
in the {\em transverse plane} $\vec x_\perp $ for each nucleon is given by a  typical  impact
parameter in $NN$ respective collisions. An estimate of it is
\be \Delta x_\perp\sim \sqrt{\sigma_{NN} \over \pi} \sim 1\, fm \ee
The uncertainty relation then tells us that each  nucleon gets a kick 
of magnitude $\Delta p_\perp\sim \hbar/\Delta x_\perp\sim 0.2\, GeV$.
This corresponds to the nucleon  kinetic energy 
\be  \Delta E_\perp \sim {\Delta p_\perp^2 \over 2M_N}\sim 20\, MeV\ee
and we suggest the {\em transverse temperature} $T_\perp$ should be of this order.

(The   exchanges of $longitudinal$ momenta in $NN$ collisions are much much larger,
but  they are not relevant for the distribution in the transverse plane we discuss.
Both small $T_\perp$ and huge $T_{long}$ will eventually equilibrate into common $T_0$,
but  we do know that did not happen at the collision moment. 
If they would, the state at the collision moment would have very high $T$ and would need a description in terms of quarks and gluons,
a la homogeneous CGC gluon state without nucleon correlations. We do know it is $not$ so,
or else fluctuations of higher angular harmonics would be much much smaller than what it is actually observed.)

Uncertainty in energy means that we will not deal with the ground state of the nucleus,
but some density matrix made out of excited states with $E_n < \Delta E_\perp$. An idea of how
it will look like can be made by assessing another density matrix corresponding
to $Euclidean$ time duration $\beta\sim \hbar/  \Delta E_\perp$. A periodic
motion with such ``Matsubara" time corresponds to density matrix of the system at certain
effective ``transverse temperature"
\be  T_{\perp}\sim \Delta E_\perp \sim 20\, MeV \ee
In other terms, we suggest that ``in anticipation of a collision" the nuclei are
``preheated" to such temperature.

(Some specialists in low energy nuclear reactions argued against ``preheating" idea, noting that
they always start with the ground state wave functions. Indeed, it is like this if only nonrelativistic quantum mechanics
of nucleon motion is considered. In relativistic case there are  ``thermal vacuum" quanta which get excited
{\em in between} two incoming nuclei about to be collided. There is no classical causality, and both
nuclei -- as well as the vacuum in between -- get ``preheated" to excited states before collision moment.
 The vacuum excitation  can borrow energy which is then returned after the collision. Furthermore,
 because of relativistic 
time delation, this time in the CM (collider) frame is increased by relativistic factor $\gamma_{CM}$. The analogy of such process to DGLAP perturbative evolution in QCD,
from the ground state to a multi-parton states with near-maximal entropy was suggested to us by D.Kharzeev in a discussion. )

Another point of discussion is why one may assume that all excited state contribute to the virtual state about equally,
allowing us to use the largest entropy argument. Thermal description of excited states goes in fact as far
 in history as Bohr's ``compound nuclei", in which also states with energy $O(10\, MeV)$ are used. However,
 we apply this description only to states from particular ``excitation trees", so its accuracy can be questioned.

\section{Classical distributions, quantum path integrals  and  semiclassical ``fluctons" } 
Let us start from the simplest proposal we have: to use  the ``deformation potentials" $E(\beta_2,\gamma...)$  
calculated by nuclear structure specialists in $classical$ Boltzmann distribution
\be P( \beta_2,\gamma..)\sim exp[-{E(\beta_2,\gamma...) \over T_\perp } ]
\ee
in defining the nuclear shape distribution. (Rather than picking up the value of shape coordinates at the 
potential minimum). Presence of two or more minima are not in this case a problem , nor is it existence of extended 
flat regions with about the same energy. 

Of course, this proposal in fact corresponds to the high-$T$ limit. 
For a general case one should use more complicated (but well developed)
computational tools  for evaluation of thermal density matrices known 
in many different branches of physics, especially in condense matter and nuclear physics. 

The density matrix with thermal weights, defined in (\ref{eq:denmatr}), is the probability $P(x_0)$ to find a system
with a particular value  $x_0$ of one coordinate.
The foundation of the method is the Feynman's path integral representation of the density matrix analytically continued to imaginary (Euclidean) time, defined as a periodic variable with period $\beta=\hbar/T$. \be
P(x_0; \beta) = {\cal N} \int_{x(0)=x_0}^{x(\beta)=x_0} Dx(\tau) \ e^{-S_E[x(\tau)]/\hbar} \ 
\label{eq:pathint}
\ee
It should be taken over the $periodic$ paths, which start  and end at the observation point $x_0$, with the period
$matching$ the duration of the Matsubara time on the circle 
\be \beta= \frac{\hbar}{T} \ , \ee
This expression
has led to multiple applications, perturbative (using Feynman diagrams) or numerical (e.g. lattice gauge theory).

At the semiclassical level, the theory is based on a classical (minimal action) periodic path,
which extends from some arbitrary point $x_0$ to the ``classical vacuum", the minimum of the potential, and return. This path has been introduced in~\cite{Shuryak:1987tr} and was named ``flucton'' (see also the lectures~\cite{Shuryak:2018fjr}).

In~\cite{Escobar-Ruiz:2016aqv} this version of the semiclassical approach was applied
for quantum-mechanical example at zero temperature.  This,  as well as
subsequent paper \cite{Escobar-Ruiz:2017uhx}, was aimed at 
developing higher order corrections in the semiclassical series, with
the one- and two-loop  quantum corrections explicitly calculated, by 
standard Feynman diagram methods for a number of quantum-mechanical problems.
These results were re-derived in~\cite{Shuryak:2018zji} from generalized Bloch equation.

Applications of the ``flucton'' method to multi-dimensional quantum systems at finite-temperature 
has been developed in \cite{Shuryak:2019ikv}, which we briefly explain here.

\begin{figure}[htbp]
\begin{center}
\includegraphics[width=7cm]{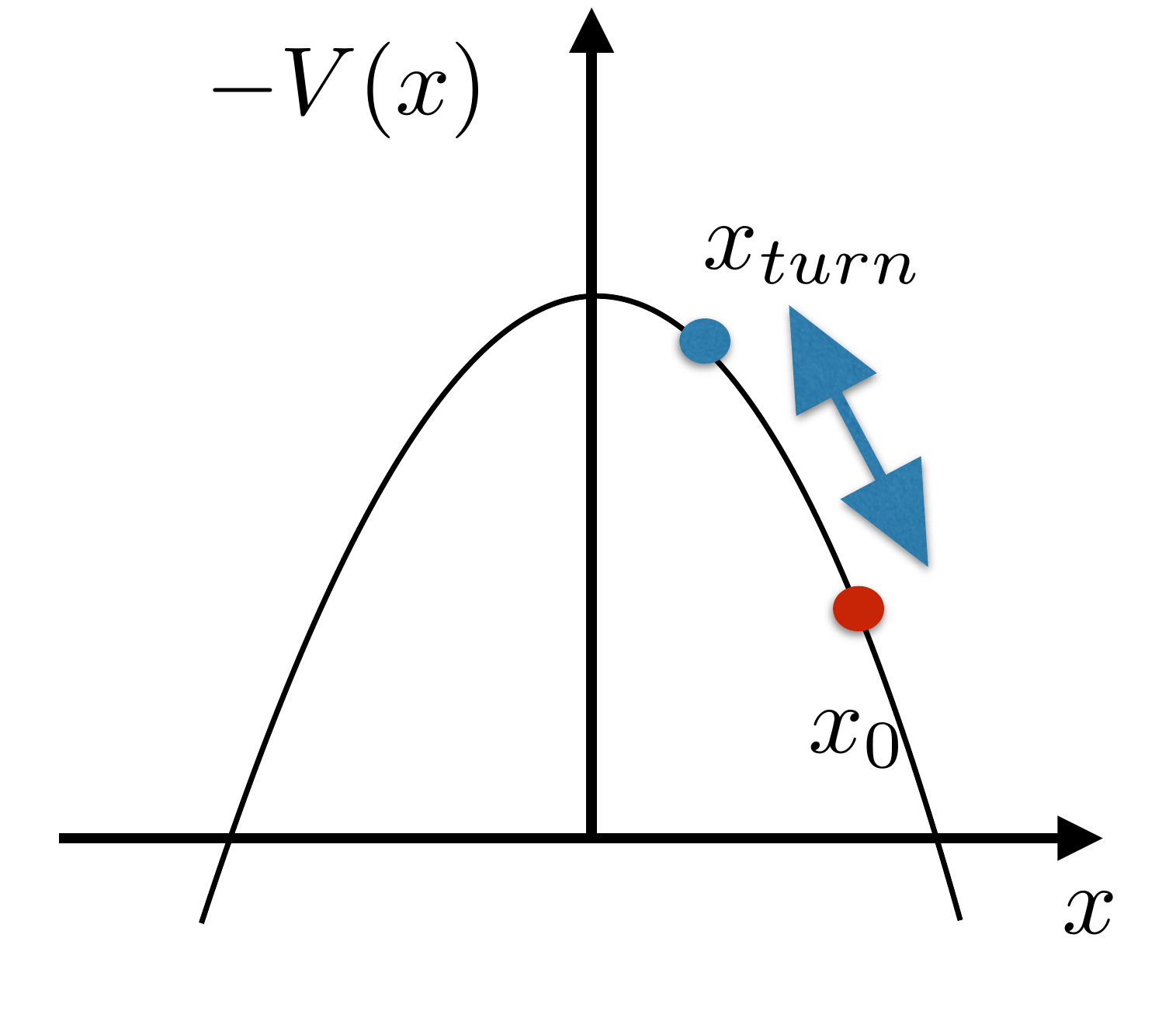}
\includegraphics[width=7cm]{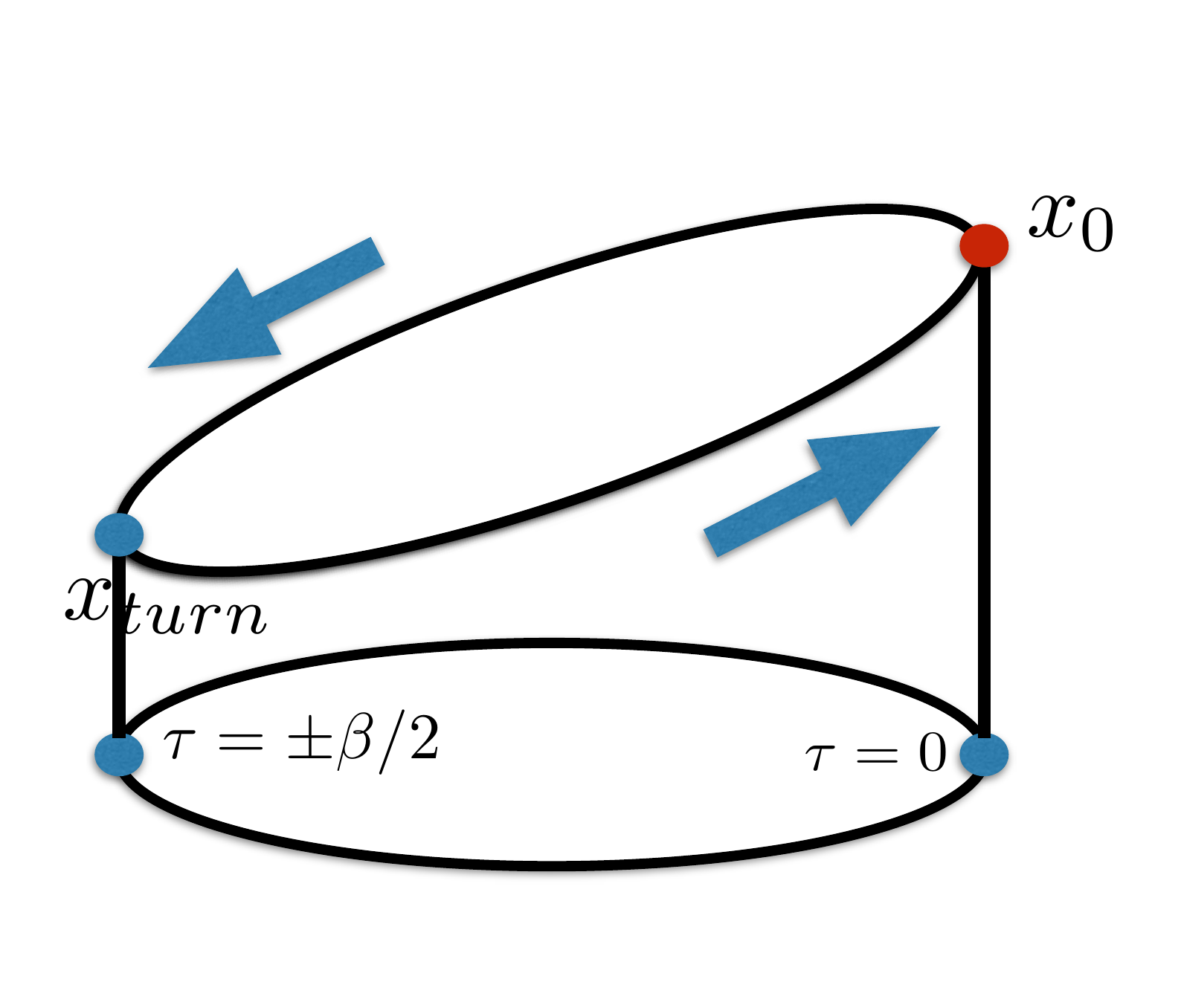}
\caption{Two sketches explaining properties of the flucton classical paths. The upper one shows the (flipped) potential $-V(x)$ versus its coordinate. The needed  path
starts from arbitrary observation point $x_0$ (red dot), goes uphill, turns back at the turning point $x_{turn}$ (blue dot), and returns to $x_0$  during the required period $\beta=\hbar/T$ in imaginary time.  
The lower plot illustrates the same path as a function of Euclidean time $\tau$ defined on a ``Matsubara circle'' with circumference $\beta$. }
\label{fig_fl}
\end{center}
\end{figure}

The ``flucton'' paths are classical solutions of the equations of motion in imaginary time (that is for a particle with Euclidean Lagrangian ${\cal L}_E$ subjected to the periodic boundary condition $x(0)=x(\beta)=x_0$. Fluctons have minimal action $S_{flucton}(x_0)$ and therefore, they dominate the path integral (\ref{eq:pathint}), provided that $S_E \gg \hbar$,
and 
\be \label{eq:dens_fluct}
P(x_0; \beta)\sim exp \big(-S_{flucton}(x_0)\big) \ee
 This definition works for both $T=0$ and $T\neq0$, and works for multidimesional systems. 
 
 The Euclidean time has $i$ and thus momentum is imaginary and kinetic energy flips sign.
 It is more convenient to flip sign of the $potential$ energy $V\rightarrow -V$ in the Lagrangian and
 EOM. Then the potential energy minima become maxima. In Fig.\ref{fig_fl} we provide two sketches explaining how these classical paths look like. At zero temperature, because in Euclidean time the potential is inverted, the particle is ``sliding'' from the maximum at $x=0$ to $x=\pm \infty$. Most of the previous applications were at $T=0$ ($\beta=\infty$) and the slide was always started from the maximum, at zero energy.
At nonzero $T$ such slides also start with zero velocity but from a certain ``turning point'' $x_{\textrm{turn}}$ and proceed toward the observational point $x_0$.

The nuclear potentials as a function of collective deformation parameters can be approximated
by some anharmonic oscillators, or perhaps sometimes even the harmonic ones.
Application of the method for harmonic and anharmonic oscillators are described in detail
in \cite{Shuryak:2019ikv}, in particular it was demonstrated that 
for the latter the density matrix calculated from
(\ref{eq:denmatr}) via sum over $O(100)$ states and via classical flucton agree very well.
For harmonic oscillator the result is analytic
\be \label{eqn_osc}
  P(x_0;\beta) \sim \exp\left[- \frac{m \omega x_0^2}{\coth (\frac{\beta \omega}{2})}\right] \ .
\ee
with the exponent corresponding to classical ``flucton" path
\be 
x_{\textrm{fl}} = x_0 \frac{e^{(\beta-|\tau|)\omega}+e^{|\tau|\omega}}{e^{\beta \omega}+1} \quad , \quad \tau \in [-\beta/2,\beta/2] \ . \label{eq:flucHO}
\ee
Note that at high $T\gg \omega$ the exponent becomes $m\omega^2  x_0^2/2T=V(x_0)/T$
corresponding to classical Boltzmann factor. In terms of flucton path this limit correspond
to the case when particle does not move at all. 

Let us now proceed to illustrate a nontrivial problem, the anharmonic oscillator, 
more relevant to generic potentials with a minimum. It is defined by 
\be S_E [x(\tau)]=\oint d\tau \left( \frac{\dot x^2}{2} + \frac{x^2}{2} + \frac{g}{2} x^4 \right) \ . \ee 
The tactics used in the previous example are not easy to implement: in particular,
the period condition defining the energy $E$ needs to be solved numerically
for each value of the $x_0$. Furthermore, using energy conservation leads naturally to
$\tau(x)$ representation of the path, rather than the conventional $x(\tau)$.
After trying several strategies we concluded that the simplest way to solve the problem is: 
\begin{itemize}
 \item [(i)] solve numerically the second-order equation of motion, 
 \be \ddot{x} = \frac{\partial V(x)}{\partial x} = x + 2gx^3 \ , \ee
 starting not from the 
observation point $x_0$ but from the turning point $x_{\textrm{turn}}$ at $\tau=-\beta/2$. This is easier because the velocity vanishes at this point, and a numerical solver can readily be used;
 \item [(ii)] follow the solution for half period  $\beta/2$ and thus
find the location of  $x_0=x(\tau=0)$;
 \item [(iii)] calculate the corresponding action and double it, to account for the other half period $\tau\in (0,\beta/2)$.
\end{itemize}

Notice that this method provides $x_0$ as an \textit{ output} after solving the equations of motion with initial conditions $x(-\beta/2)=x_{\textrm{turn}}$ and $\dot{x}(-\beta/2)=0$. One could also tweak a bit the method to use $x_0$ it as an \textit{ input} by using the constraints $x(0)=x_{0}$ and $\dot{x}(-\beta/2)=0$.

\begin{figure}[htp]
\begin{center}
\includegraphics[width=7cm]{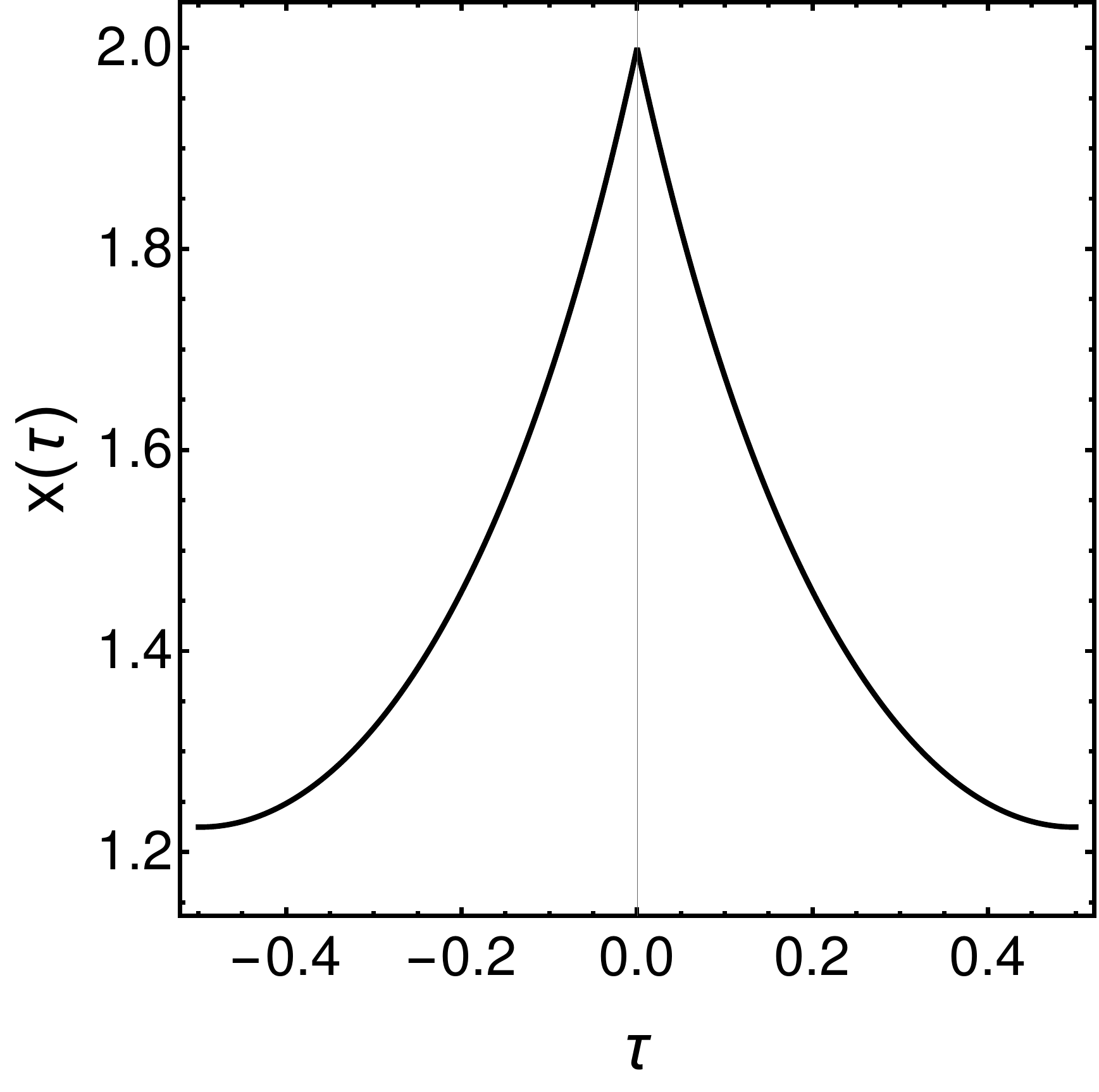}
\caption{Flucton path for the anharmonic oscillator with $g=1$ and $T=1$ (in units of the mass), for the observation point $x_0=2$. Notice that, as expected, $\tau \in (-\beta/2,\beta/2)$ with $\beta=1/T=1$ and $x(\tau=0)=x_0$.}
\label{fig:flucAHO}
\end{center}
\end{figure}

In Fig.~\ref{fig:flucAHO} we show the numerical solution of the flucton path for the anharmonic oscillator with $g=1$ and $T=1$ (in units of the mass). We choose the observation point $x_0=2$, which is reached as expected, at $\tau=0$ (cf. Fig.~\ref{fig_fl}). The flucton is periodic in $\tau$ with period $\beta=1/T$.

\begin{figure}[htp]
\begin{center}
\includegraphics[width=7cm]{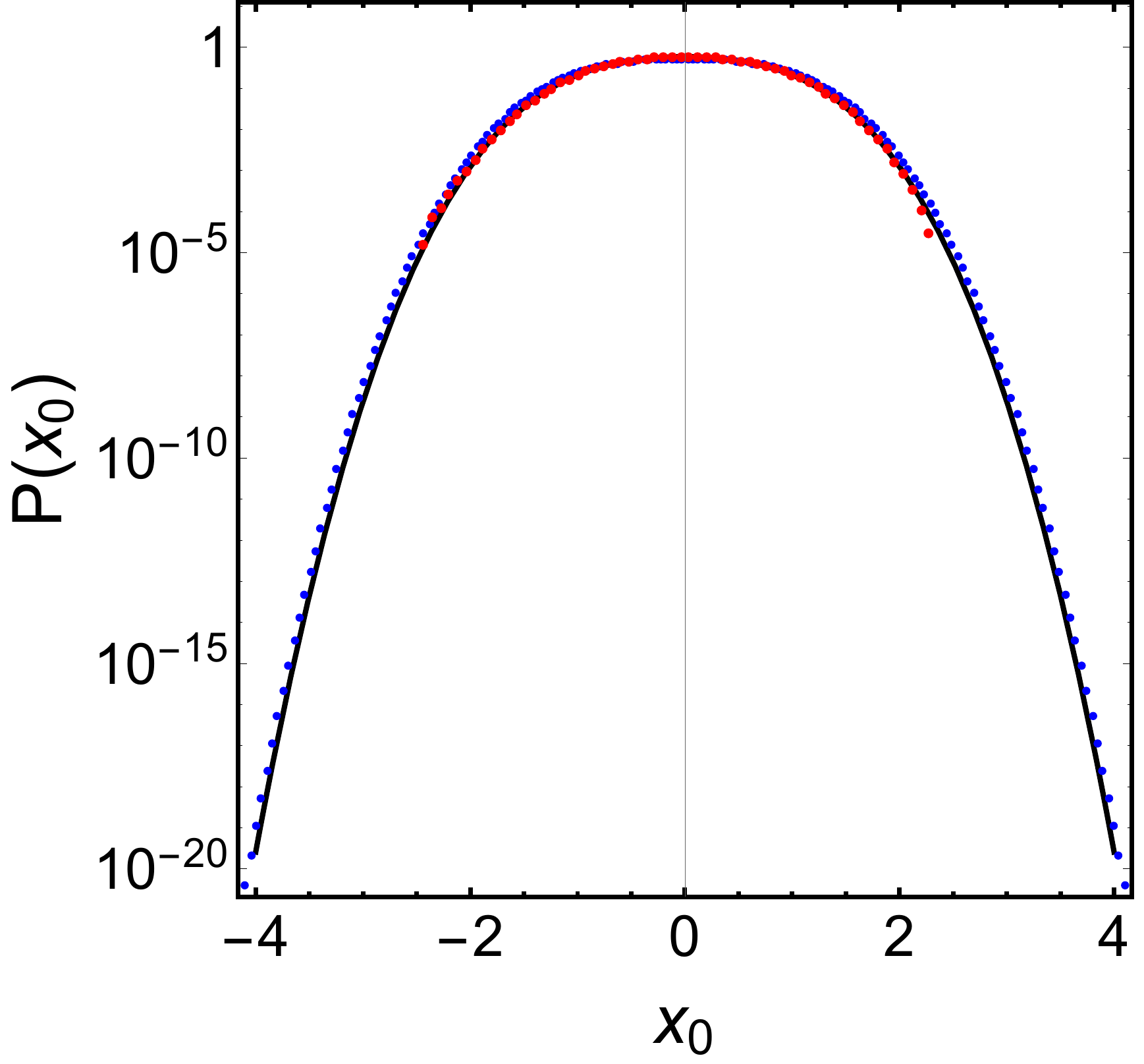}
\includegraphics[width=7cm]{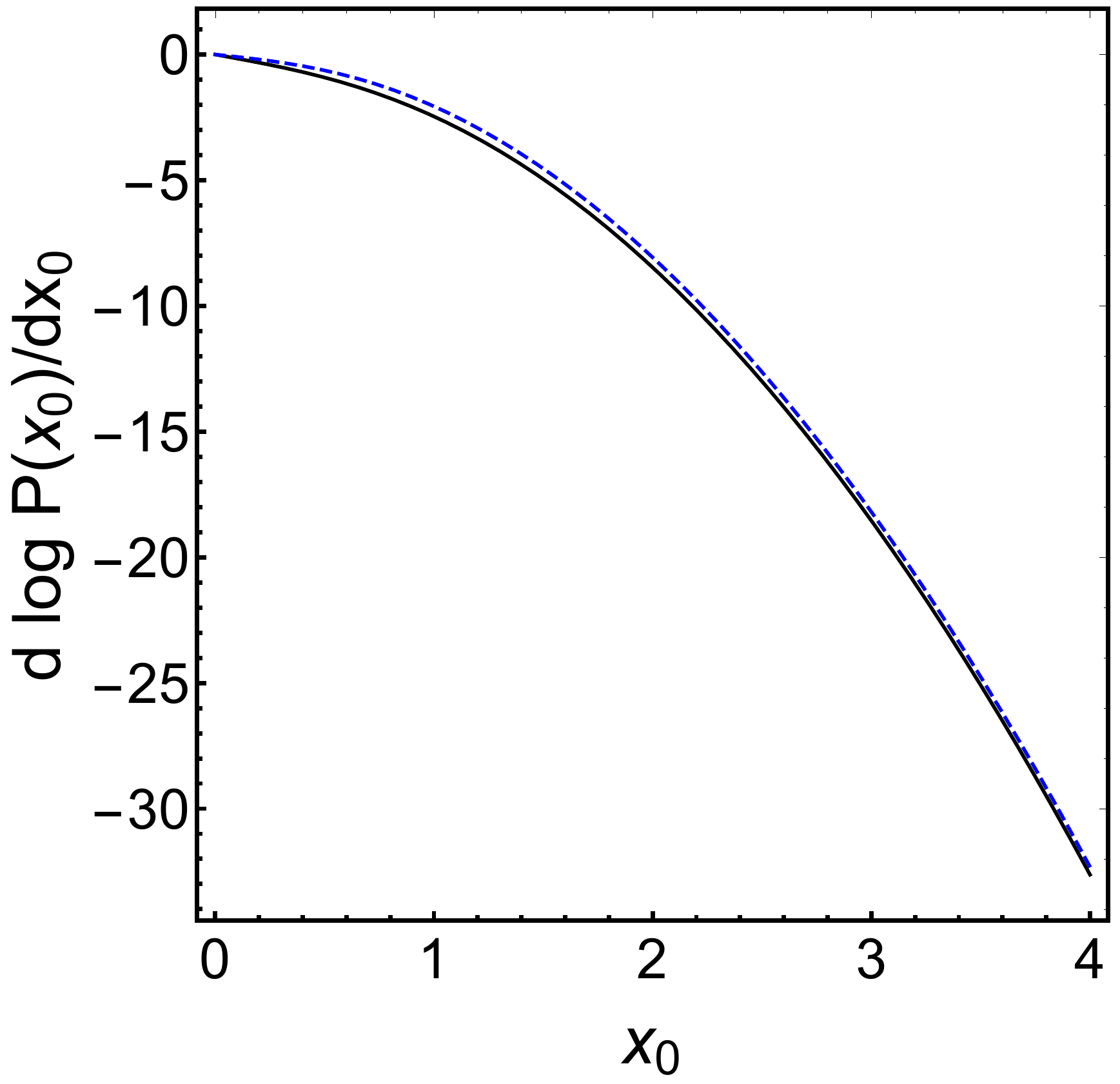}
\caption{Left panel: Density matrix $P(x_0)$ vs $x_0$ for anharmonic oscillator with the coupling  $g=1$,
at temperature $T=1$, calculated via the definition of summing Boltzmann-weighted  states (line) and
the flucton method (points). The line is based on 60 lowest state wave functions found
numerically. Right panel: Comparison of the logarithmic derivative of the density matrix of the upper panel.}
\label{fig_anharmonic}
\end{center}
\end{figure}

Here we present the upper panel of Fig.~\ref{fig_anharmonic} comparing the summation over 60 squared wave functions, and Boltzmann weighted (solid line),  with the result of the flucton method (points) at $T=1$ (in units of the mass). The coupling is set to $g=1$.  For additional comparison we also got numerical results of a path integral Monte Carlo calculation with the same parameters (not shown).

As a semiclassical approach one expects that the flucton solution works better when the action is large, i.e. for large values of $x_0$. However, one observes that the flucton systematically overestimates the solution based on the Schr\"odinger solution. Part of the discrepancy comes from normalization issues as described in~\cite{Escobar-Ruiz:2017uhx}. To remove those it is enough to compare the logarithmic derivative of the density matrix $d \log P(x_0)/dx_0$. In the bottom panel of Fig.~\ref{fig_anharmonic} we show the logarithmic derivative of the density matrix in linear scale. While the agreement is nearly perfect, a small difference can still be detected. We ascribe it to the ``loop" corrections to the thermal flucton solution~\cite{Escobar-Ruiz:2017uhx}.

(As we already mentioned, the actual application on which \cite{Shuryak:2019ikv} was focused was multi-nucleon correlations
at freezeout stage of heavy ion collisions, important for light nuclei production. 
This problem is multi-dimensional and thus one by necessity needs to
define one {\em collective variable} hyperdistance $\rho$ and study thermal density matrix $P(\rho)$.
Derivation of ``flucton" path  was based on corresponding Schreodinger equation in 9 dimensions.
The method was checked later in \cite{DeMartini:2020anq} where finite-T path integral was done numerically. )

\section{What excitation spectra of both ``isobar" nuclei can tell us about their
density matrices  }
We now return to particular nuclei  $^{96}_{40}Zr$ and  $^{96}_{44}Ru$ and note that already standard
shell model calculations show that there should be significant difference between them (see Appendix).
The $pp$ and $nn$ pairs get strongly correlated by Cooper pairing. Since there are several such pairs,
their states are not simple, and this is what nuclear structure professionals calculate.

The spectroscopy of excited states of two nuclides in question provides key information about their structure.
Before we go to specifics, let me note that
 experimentally it is followed till around nucleon separation energy or $O(10\, MeV)$.
Since
 our estimated $T_{\perp}$ is unfortunately higher,
 we will not yet have full set of excitations
needed to calculate the thermal density matrix from them. Yet we do have enough excitations
to understand what are the main excitations types of both nuclei,
whether they are ``rotors" or "oscillators"  and with what parameters.

\subsection{$^{96}_{40}Zr$, its configurations and ``excitation trees"}
One family of (collectvized) particle-hole bound states
are known as {\em nuclear phonons}. In first approximation their effective
Hamiltonian is that of harmonic oscillator, and the lowest states are approximately equidistant.
The quantum numbers of a ``phonon" depend on those of particle-holes, and
those of n-phonon states can be deduced from those using standard rules of summed angular momenta. For example, most typical quadrupole oscillation phonons have $J^P=2^+$,
two-phonon states around twice excitations are with $J^P=4^+,2^+,0^+$, etc. 
More accurate description is provided by anharmonic oscillators for 
``interacting bosons model"  IBM, for recent discussion of Zr isotopes in it and 
general references see \cite{Gavrielov:2021vck}.

One important concept is that nuclei can  be thought of in terms of several {\em coexisting
configurations}. Furthermore, each configuration has its own {\em excitation tree} (also called a ``band").
Since  transitions between states are mainly confined inside each, these trees are
relatively distinct experimentally (see below).

In the particular case of  $^{96}_{40}Zr$ configuration A correspond to closed proton 
sub-shell and only $nn$ pairs, while the configuration B contains two proton 
excitation (from below to above sub-shell gap) with a $2p-2h$ state, etc. Each
of them have their effective Hamiltonians $H_A, H_B...$, with relatively small but
nonzero mixing terms $H_{AB},...$ (we will further ignore). The IBM approach
is to formulate Hamiltonians not in terms of quasiparticle pairs but in terms of 
scalar and quadrupole ``phonons". Their numbers are defined as
\be \hat n_s+\hat n_d=s^+s+\sum_\mu d_\mu^+ d_\mu \ee
(Microscopic derivation of IBM relates number of phonons to number of quasiparticle pairs,
but we will not discuss that.)

The other important concept is that of {\em dynamical symmetries}. Unlike the usual symmetries,
it does not imply certain operators to commute with Hamiltonian, but that a number of operators
form a closed algebra, and thus states can be calculated algebraically, using representations
of the corresponding groups. Since there are 1+5=6 phonon operators, the largest group is $U(6)$, which in particular case can be reduced to its subgroups ($U(5), SU(3), SO(6)$ etc).
Related to that
is a concept of {\em collective motion paradigms},
which correspond to such dynamical symmetries. The simplest is spherical
vibrations $[ U(5)]$, or axially symmetric $[ SU(3)]$, or $\gamma$-soft deformed rotor $[ SO(6)]$, etc. Geometrical interpretation of states obtained can be visualized by {\em coherent
states} with certain parameters, such as quadrupole shape parameters $(\beta, \gamma)$
related to the following creation operator
\be b^+={1 \over \sqrt{1+\beta^2}}\big( s^++ \beta cos(\gamma) d_0^+ + \beta sin(\gamma)
(d_2^++d_{-2}^+)/\sqrt{2}
\big) \ee
The IBM Hamiltonians are made of quadratic part in $s,d$ operators and quartic
one, typically in form of quadrupole-quadrupole form, with quadrupole quadratic in  $s,d$.
The Hamiltonian averaged over these states defines the ``energy profile"
\be E(\beta,\gamma)=\langle \beta,\gamma | \hat H | \beta,\gamma \rangle \ee 
describing quantum motion in terms of the corresponding collective variables.

In the chart of nuclides $(Z,N)$ there exist multiple domains in which excitation trees
have the same symmetry, and effective Hamiltonian just display smooth change of
parameters. They are separated by lines of 
{\em ``mini phase transitions"}. We put these word into parenthesis for few reasons.
First of all, these transitions happen  for each "excitation trees" individually. Second,
they indicate excitations of just several (not macroscopically large) number of pairs:
therefore they would only be observed by high accuracy data.
And, finally, since $(Z,N)$ changed in a discrete manner (by two protons or neutrons,
for even-even nuclei) there is no true critical points or singularities, but just jumps from one
phase to another.

\begin{widetext}
\begin{figure}[t]
\begin{center}
\includegraphics[width=17cm]{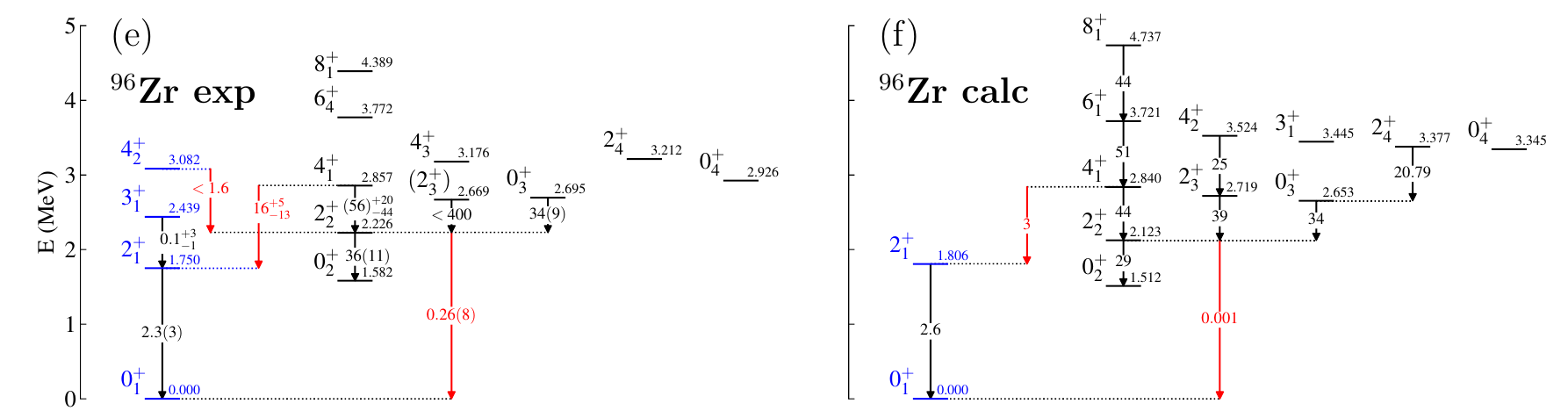}
\caption{Blue (left) and black (right) are states corresponding to ``excitation trees" growing from configurations A and B,
respectively. }
\label{fig_Zr_spectra}
\end{center}
\end{figure}
\end{widetext}

\begin{figure}[h!]
\begin{center}
\includegraphics[width=5cm]{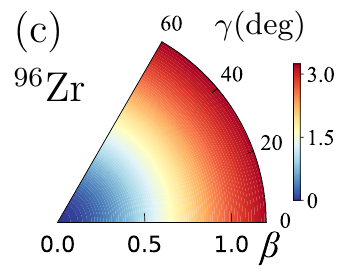}
\includegraphics[width=5cm]{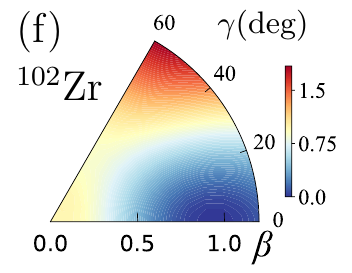}
\includegraphics[width=5cm]{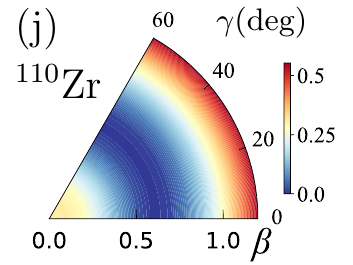}
\caption{Contour plots of $E(\beta,\gamma) $ for three $Zr$ itsotopes, displaying different
symmetries}
\label{fig_beta_gamma_Zr}
\end{center}
\end{figure}

Let us show how it looks in practice, for particular nucleus in question.   The experimental and calculated parts of the spectra, from \cite{Gavrielov:2021vck}, are shown in Fig.\ref{fig_Zr_spectra}. Focusing on configuration B excitation tree (black, right) one observes
typical set of states of a (slightly anharmonic) oscillator, with $2^+$ phonon state, $4^+,2^+,0^+$ two phonons, up to three phonons states. The ratio of their energies to that of a single phonon
are indeed close to 2, 3 etc., confirming vibrational interpretation of the tree.
three phonons etc.  
The corresponding picture of $E(\beta,\gamma) $ is given in Fig.\ref{fig_beta_gamma_Zr},
for three $Zr$ isotopes. As one can see, they correspond to qualitatively different
"phases" of configuration B. The one we focus on, $^{96}Zr$ 
has a potential with a single minimum at the origin, corresponding to basic spherical shape.
Its potential seems to be independent on angle $\gamma$.

But  already the  isotope $^{102}Zr$ (with 3 extra $n$ pairs.) show a completely different
potential: now the minimum is at large $\beta$ and zero $\gamma$. Adding 4 more neutron pairs  to $^{110}Zr$ we again find that another ``mini phase transition" line was crossed, since
the shape of the effective potential gets qualitatively different once again.

Now we return to our main problem, evaluation of the density matrix. If the collective motion
is described by a harmonic oscillator, the probability to find (configuration B) nucleus 
with particular $\beta,\gamma$ is then Gaussian (\ref{eqn_osc}). Furthermore,
when $T>>\omega\approx 2.226-1.582=.644\, MeV$, thermal density matrix
should be given  just  by the classical Boltzmann factor
\be  P(\beta,\gamma)\sim exp\big[ - {E(\beta,\gamma) \over T}\big]
\ee

\subsection{$^{96}_{44}Ru$: deformations  and rotations}
We now focus on the second nuclide used in STAR experiment.
Reducing $^{96}_{44}Ru$ problem  to four pairs,
1 $nn$ pair and 3 $pp$ ones, may appear a simpler  problem, yet  there are
4 pairs of $\theta,\phi$ variables. Doing quantum mechanics in 8-2=6 dimensions
(global orientation obviously cannot matter) is still  not easy.

\begin{figure}[h!]
\begin{center}
\includegraphics[width=14cm]{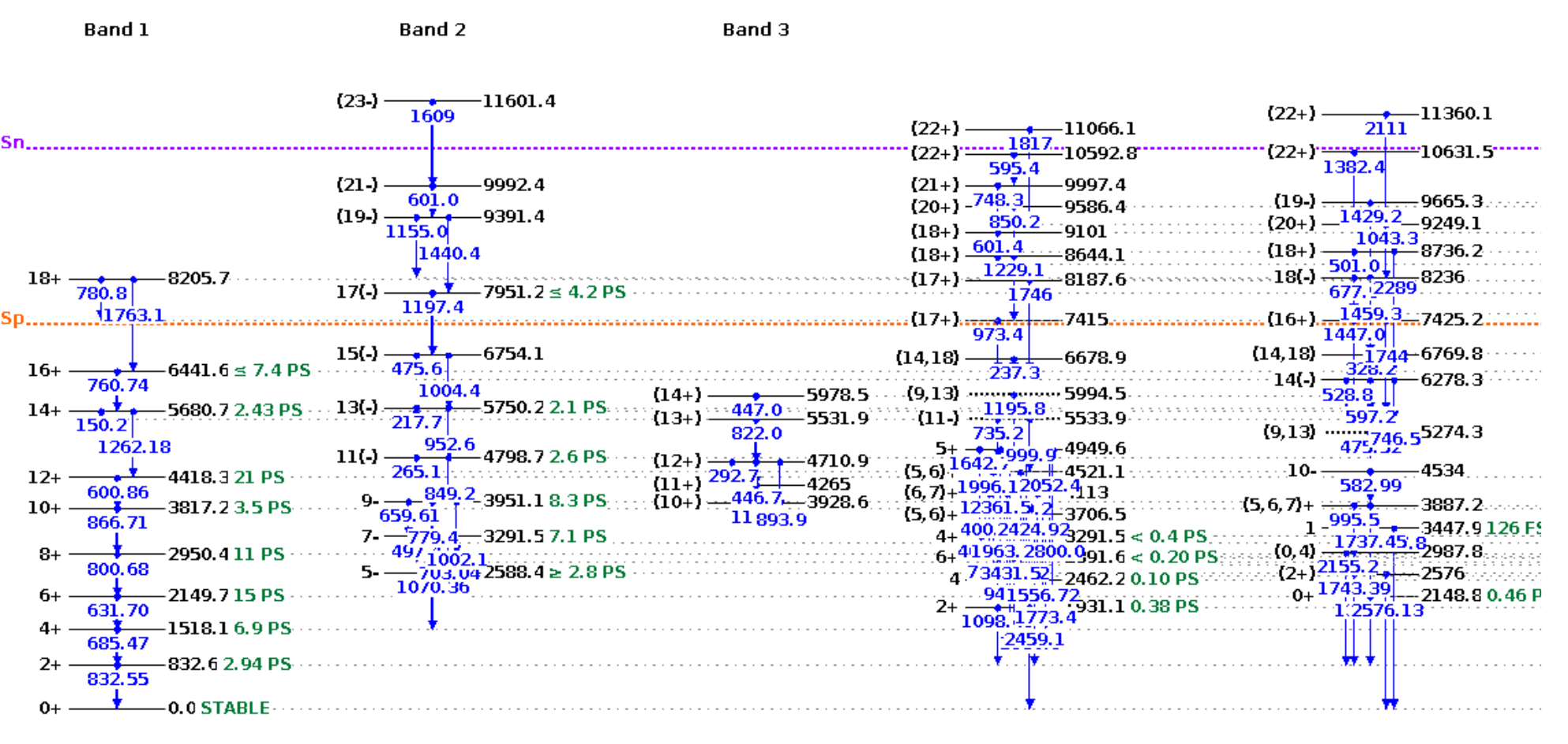}
\caption{Excitation levels of $^{96}Ru$ (replotted from the BNL webpage of nuclear excitations)}
\label{fig_Ru_levels}
\end{center}
\end{figure}

Fortunately, a lot of information is available about the excitations, see Fig.\ref{fig_Ru_levels}. Clear
separation into excitation trees or five ``bands"  are shown. 

The first one is a set of states with $J^P=0^+,2^+....18^+$, a typical {\em rotational band}.
Since spherical nucleus cannot be rotated, we learned that this band
corresponds to a  deformed but axially symmetric configuration. 

Two ways how information on the band can be used. We define
$J$-dependent moment of inertia and rotational frequency by
\be  I_J=  {J (J+1) \over 2 E_J},   \,\,\,\,\omega_J= {E_{J+2}-E_{J} \over 2} \ee
and get for the former ($GeV^{-1}$)
$$ I_{J=1..8}=3603.17, 6587.18, 9768.8, 12201.7, 14408.5, 17653.8, 18485.9, 21112.8$$
Here we see that nuclei are ``flexible" (not rigid), with
momentum of inertia (and thus deformation) growing with $J$. It  remains significantly smaller than the moment of inertia for
 ``solid state sphere rotation", which for a sphere is
$$ I_{solid}=(2/5)*M_{tot}R^2\approx 32400. GeV^{-1} $$
Therefore, only a part of nuclear matter is actually rotating (which  is known since
1950's).  Again: defining deformation at the collision moment, one has to specify how many states
are included in the wave package, or how much preheated the nuclei actually are.

At the other hand, consider  one compact Cooper pair sitting at the equator: it will add
to moment of inertia an amount
$$I_{pair}= 2MR^2 \approx 1700 GeV^{-1} $$
which is smaller than the observed $I_J$ values. But, of course, there are four
Cooper pairs sitting somewhere on a sphere, and the observed values can
correspond to some particular arrangements of those. Clearly, as $J$ grows,
the pairs become unpaired by centrifugal force and become ``normal", thus growing
$I_J$. 

Looking at $\omega_J$ from the rotational band one finds that it is nearly constant.
This indicates that all excitations rotate with about the same rotational frequency,
and all increase in $J$ is dues to increase in momentum of inertia. The ``unpairing"
of Cooper pairs is not a sharp transition, like observed in heavier nuclei, but gradual
unpairing of quasiparticles. 

Let us now discuss the second band (tree of similar states). All of them are $P=-$, so clearly they are not
axially symmetric. The root of this tree is $5^-$ state, which obviously cannot be described in
an IBM usual building bocks,  $0^+$ and $2^+$ phonons: some Cooper pair should be unpaired for that. Further excitations in this tree also indicates rotations. (Addition of quadrupole phonons cannot describe it since it would generate many more states which are not there.) 

Now we learned an important lesson: superpositions of excitations from both first trees
would generate parity-odd terms in the density matrix, e.g. $3^-$ or pear-like shapes.
If so, one may expect triangular flows in STAR experiment with this nuclide, as indeed was found.

\section{Conslusions}
High accuracy of STAR data allows us a rare opportunity to test at entirely new level our understanding of nuclear shapes, via comparison of the multiplcity distribution, as well as elliptic and triangular flows. There
are several studies using density functional or ``neutron skin" data to argue that, contrary to Coulomb effect,
neutron-rich isobar has a larger radius. We show in Appendix that  one comes
to a very similar conclusion using standard  shell model states. 

The central idea is that the state of nuclei at the collision moment is $not$ described by its ground state
but a certain wave package made up of $many$ excited states. Arguments based on density of state
(maximal entropy) suggest to describe those as a {\em thermal state} with some temperature $T_\perp$.
The ``intrinsic deformations" of nuclei can then be described using ``potential energies" already
calculated by nuclear structure practitioners. 

If temperature is high enough the distribution over collective variables can be described just by 
Boltzmann distribution with those potentials. More accurately, it can be described by 
{\em semiclassical  flucton method } at nonzero temperature, which correctly includes both quantum and
thermal fluctuations. We have shown this method to be very accurate for anharmonic oscillators,
of the type to be relevant to the fluctuations in nuclear deformation parameters $\beta_2,\beta_3,\gamma$.

In this note we also focused on the  ``excitation trees" corresponding to coexisting configurations
of the corresponding nuclei. It is known, and demonstrated for B configuration of $Zr$ series in \cite{Gavrielov:2021vck},
that such trees undergo ``mini phase transitions" along certain lines on the
nuclide chart $(N,Z)$, at which the nature of collective excitations changes qualitatively.
Crossing such lines would induce jumps in many observables, including the 
angular moments of the density
matrix  which seeds the collective flows. 

There is no doubt that  going from $^{96}_{40}Zr$ to  $^{96}_{44}Ru$ such lines are crossed,
as the former is basically a spherical nucleus with phonon-like excitations, while the latter
is a deformed one with well developed rotational bands. That is why the measurements
shown in Fig.1 had shown deviations from 1 by as much as $10\%$, dwarfing
CME and other Z-related effects. 
If this type of isobar pair experiments will be planned in the future, one needs
to check whether both nuclei are $not$ separated by mini phase transition lines.

One final thought deals with methods to measure nuclear charge distributions using ultraperipheral 
$e^+e^-$ pairs. The ``preheating" idea suggest that sizes of nuclei about to collide with another nucleus are
a bit larger than it is for the same nucleus at rest (or in EIC collisions in which collisions are with an electron/photon). 

\appendix
\section{Quasiparticles in nuclear shell model}
\begin{figure}[h!]
\begin{center}
\includegraphics[width=6cm]{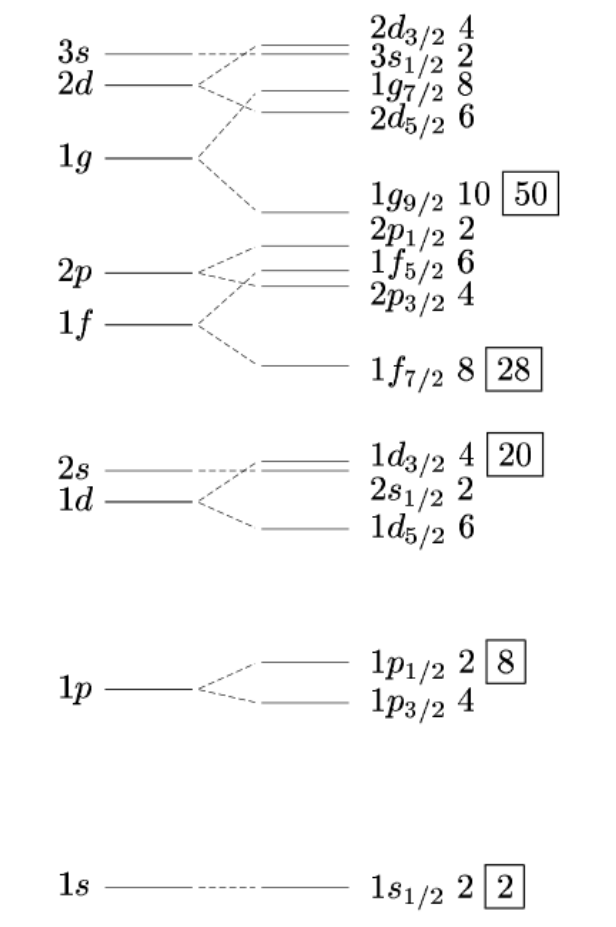}
\caption{Shell model levels from textbooks.}
\label{fig_shell_model_plot}
\end{center}
\end{figure}

The shell model single-nucleon states, calculated in a collective nuclear potentials, are filled in the order prescribed
by one-nucleon energies, as  shown in a textbook Fig.\ref{fig_shell_model_plot}.

As it follows, 50 is a ``magic number", and the double-magic $^{100}Sn$ should be a
nice spherical nuclei with filled shells. The nuclei we are interested in
differ from it by (2 or 6) neutrons in the $2d_{5/2}$ state and (6 or 10) 
proton holes in $1g_{9/2}$ states. Note that those states have very different
radial dependence, differing not only in orbital momentum (2 versus 4) but even in principal
quantum number. 

\begin{figure}[h!]
\begin{center}
\includegraphics[width=6cm]{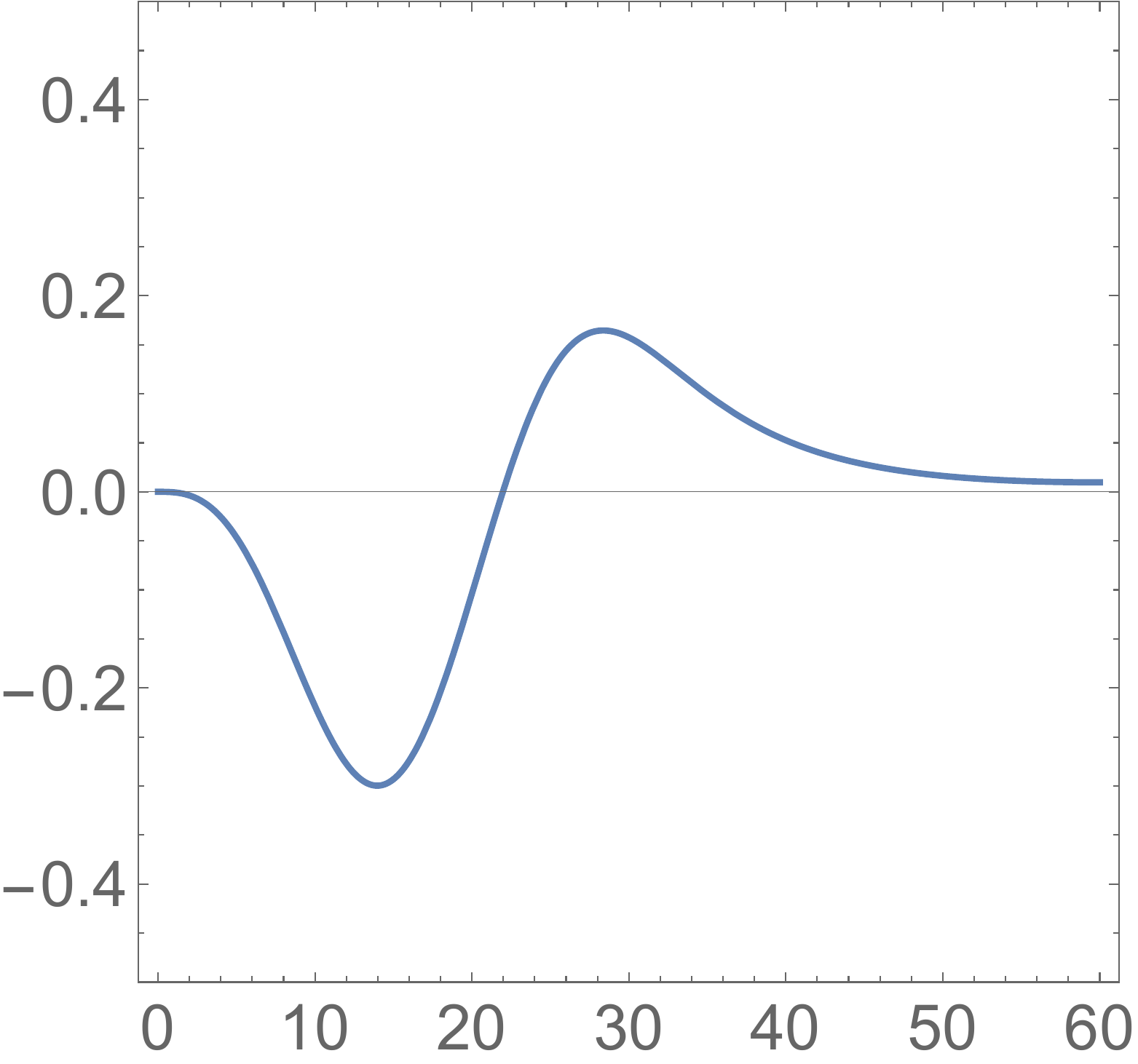}
\includegraphics[width=6cm]{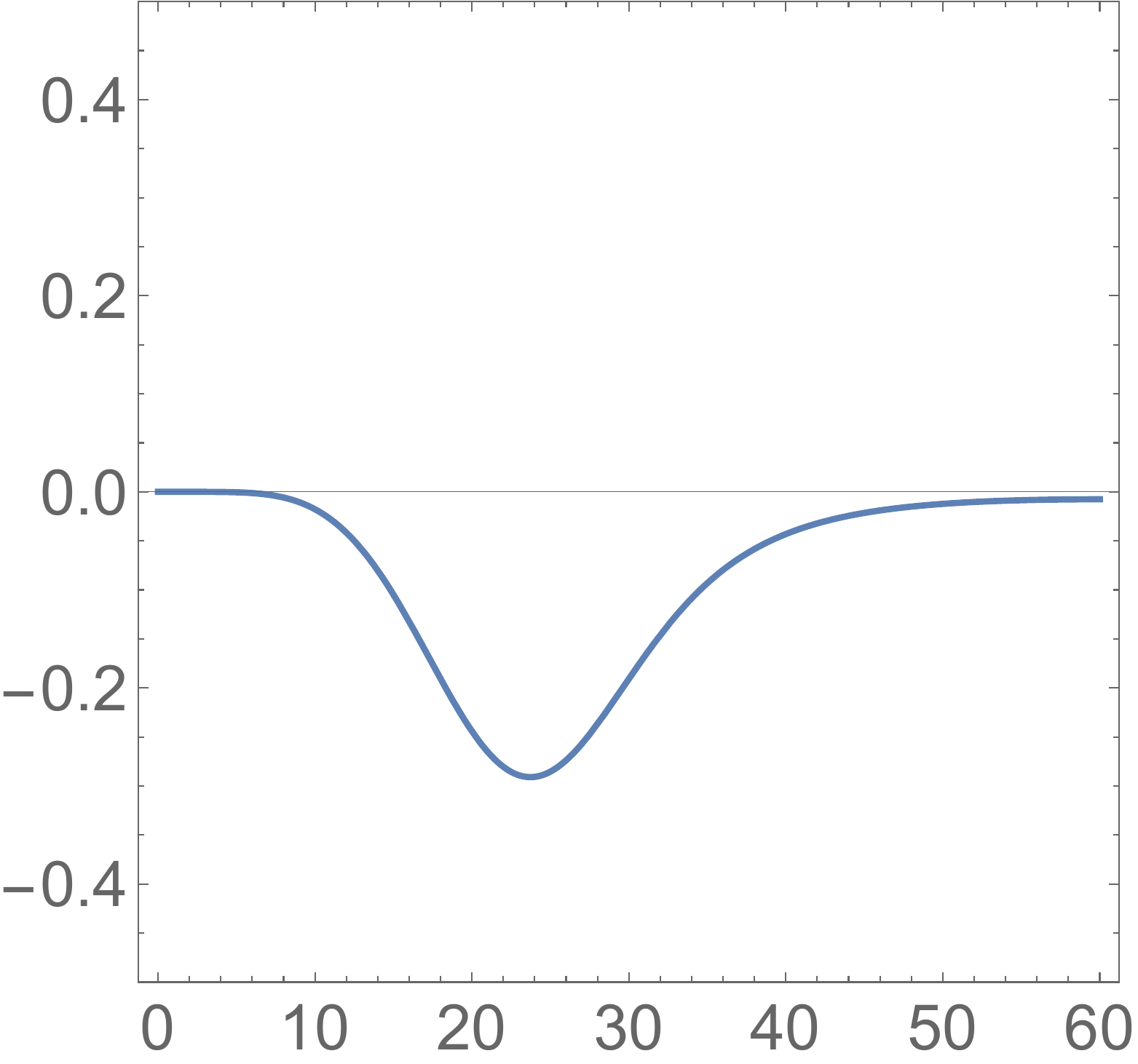}
\caption{(Unnormalized) wave functions $2d$ and $1g$}
\label{fig_psi}
\end{center}
\end{figure}

Let us calculate the corresponding wave functions. Using nuclear potential 
\be V(r) := -{V_0 \over 1 + Exp[(x - R)/a] } \ee
with $R = (1.25/0.197)*96^{1/3}, V0 = .057; a = .65/0.197$ (all  energies in $GeV$, distances in inverse $GeV$) we calculated the corresponding wave functions, see
Fig.\ref{fig_psi}.
\begin{figure}[h!]
\begin{center}
\includegraphics[width=6cm]{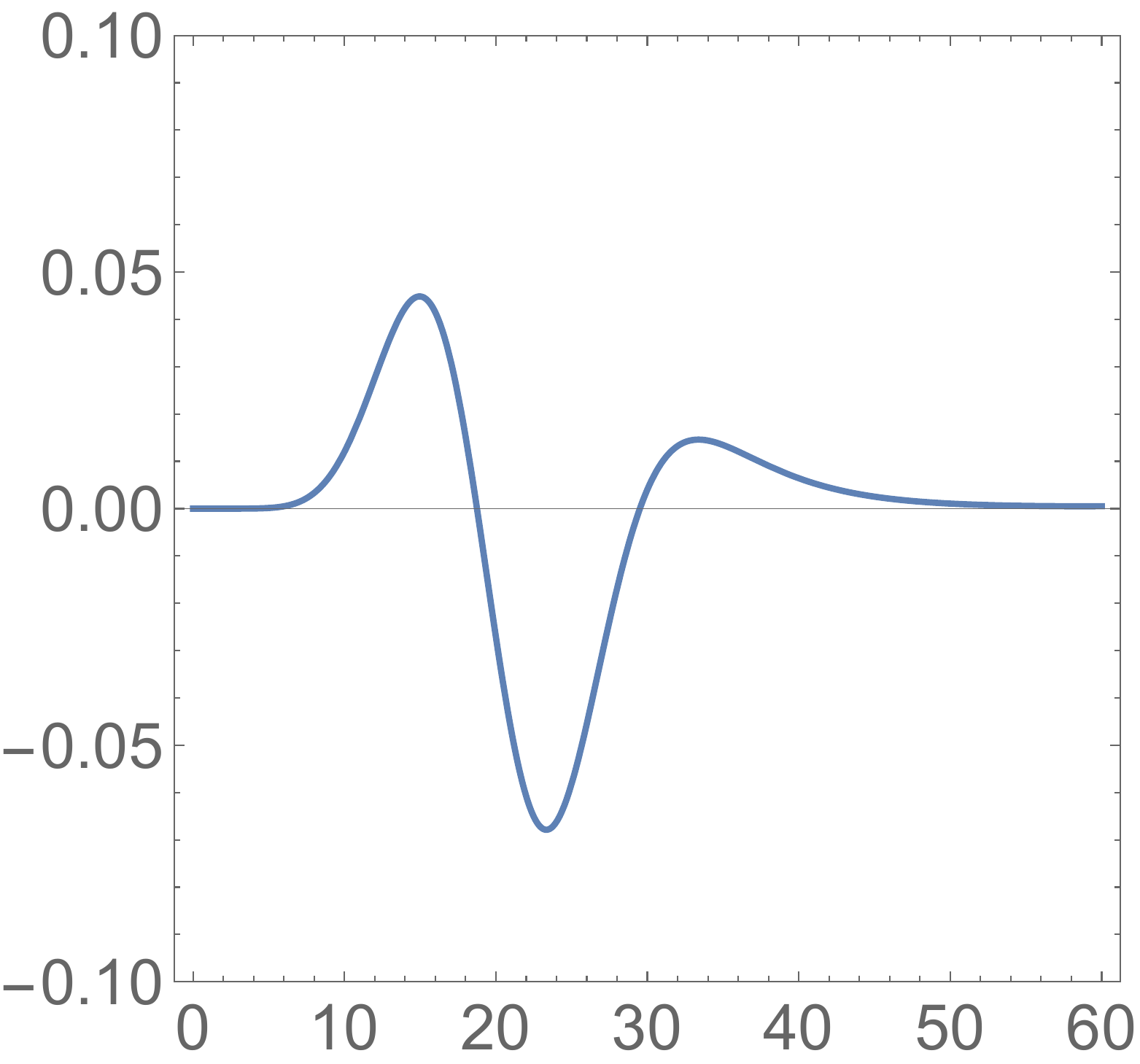}
\caption{The density ratio $Ru/Zr$ as a function of the distance ($GeV^{-1}$).}
\label{fig_radial_difference}
\end{center}
\end{figure}
Indeed, they have  very different shapes.  Note,
that the former one has a node, located exactly where the latter has a maximum. 

%
The radial dependence of densities  can be taken in the form
\ba Ru = d_{5050}(r) - 6 \psi_p^2(r) + 2 \psi_n^2(r) \nonumber \\
Zr = d_{5050(r)} - 10\psi_p^2(r) + 6  \psi_n^2(r);\ea
where the first term is a parameterization for the double-magic 50-50 nucleus. Their difference is shown in Fig.\ref{fig_radial_difference}.Note a certain excess of $n$ at large $r$: while it is qualitatively similar to a ``halo" discussed in literature,
but it is not due to manybody effects but just follows from the shapes of the single-body wave functions.

\vskip 1cm
{\bf Acknowledgements}
This work is supported by the Office of Science, U.S. Department of Energy under Contract No. DE-FG-88ER40388.
I also should thank J.Jia and other organizers of BNL workshop on the subject, which prompted me to put
these ideas on paper and added a talk at very short notice. 

\bibliography{isobars}

\end{document}